\documentstyle[preprint,pre,aps]{revtex}
\begin{document}

\draft
\title{Chiral Twisting of a Smectic-$A$ Liquid Crystal}
\author{M. S. Spector,$^1$ S. K. Prasad,$^{1,2}$
B. T .Weslowski,$^{1,3}$ R. D. Kamien,$^{4}$ J. V. Selinger,$^{1}$
B. R. Ratna,$^1$ and R. Shashidhar$^1$}
\address{
$^1$Center for Bio/Molecular Science and Engineering,
Naval Research Laboratory, Code 6950, Washington, D.C. 20375-5348 \\
$^2$ Centre for Liquid Crystal Research,
Jalahalli, Bangalore 560 013, INDIA \\
$^3$ George Mason University, Fairfax, Virginia 22030 \\
$^4$ Department of Physics and Astronomy,
University of Pennsylvania, Philadelphia, Pennsylvania 19104
}
\date{\today}
\maketitle

\begin{abstract}
Chiral twisting of the molecular orientation within the layer
of a smectic-$A$ liquid crystal has been investigated 
using circular dichroism spectroscopy.
The results indicate that a rotation of the layers away from
the alignment direction is induced by
the surface electroclinic effect.
This leads to an interfacial region where the molecular
director twists from the alignment direction until
it reaches the layer normal direction.
A theory is presented to
explain the observed field and temperature
dependence of the circular dichroism.
\end{abstract}


\narrowtext

\section{Introduction}

The addition of chirality to the molecular structure
of smectic liquid crystals leads to unique materials
properties which have allowed for the development
of fast, high-resolution display devices \cite{walba95}. 
Molecular chirality results in ferroelectricity in the
tilted smectic-$C^\ast$ liquid crystalline phase \cite{meyer75}.
In the non-tilted smectic-$A$ phase, rotational symmetry
about the molecular long axis rules out any spontaneous 
polarization in the bulk.
However, application of an electric field parallel to
the smectic layers breaks this symmetry due to the
coupling of the field to the transverse dipole moment and
leads to an induced tilt relative to the layer normal,
the electroclinic effect \cite{garoff77,lagerwall97}.
A similar symmetry-breaking effect can also arise 
due to surface interactions.
Unidirectional rubbing of a cell surface is commonly
used to align liquid crystals so that their long axes
are along the rubbing direction.
In the smectic-$A$ phase, the layer normal 
is also typically parallel to the rubbing direction resulting
in the planar or bookshelf geometry.
However, Nakagawa {\it et al.} observed that for a smectic-$A$
composed of chiral molecules, the optic axis is rotated with 
respect to the rubbing direction \cite{nakagawa88}.
This result was attributed to a polar interaction between
the liquid crystal molecules and the surface which gives
rise to a surface electroclinic effect \cite{nakagawa88,xue90}.
Subsequent studies on another material found that the rotation
angle of the optic axis can be as large as 18$^\circ$ \cite{patel91}.

Detailed studies by Chen and coworkers, using optical
second-harmonic generation, found that the molecules
anchored at the cell surface are oriented along the
rubbing direction, while the optic axis of the bulk
is rotated with respect to the rubbing direction
by an angle $\Psi$ \cite{chen92}.
Taking the optic axis to be parallel to the
layer normal in the bulk, this suggests that
direction of the layer normal is determined
by surface interactions and remains constant
as one moves from the surface to the bulk.
Thus, unlike the field induced electroclinic effect
that results in a uniform tilt of the
director over the entire thickness of the 
cell, the surface electroclinic effect is a
localized effect which is restricted to a boundary
layer near the cell surface.
This layer is much thinner than the wavelength of
visible light, since a uniform optic axis
and good extinction are observed when a thick cell is
viewed between crossed polarizers \cite{patel91}.
In the interfacial region, the molecular director
twists by $\Psi$ from the rubbing axis until it
reaches the layer
normal direction as one moves away from the surface to
the interior of the cell as illustrated in Fig.\ 1.
The thickness of the boundary layer $\xi$ is determined by the
correlation length of tilt fluctuations in the smectic-$A$ phase.
Using ellipsometry, Chen {\it et al.} found the expected
divergence of $\xi$ with decreasing temperature near the
smectic-$A$ to smectic-$C^\ast$ transition \cite{chen92}.

In this paper, we use circular dichroism (CD) spectroscopy
to directly probe the chiral twist between the 
surface and bulk states.
Circular dichroism is a sensitive measure of the twisted
orientation of molecules with an appropriate chromophore
in the chemical structure of the molecule.
Past studies have found large 
CD peaks in the ferroelectric smectic-$C^\ast$ phase and
pretransitional effects in smectic-$A$ phase when 
approaching smectic-$C^\ast$ phase \cite{li95}. 
This chiral order arises from helical variation of the
azimuthal tilt direction from layer to layer and is
observed for light propagating perpendicular,
but not parallel, to the smectic layers.
The present work represents the first observation of CD
due to the chiral order induced by the surface
electroclinic effect
and is only observed when
the light is propagating parallel to the smectic layers.
Our results show the existence of a strong twist over a
wide temperature range in the smectic-$A$ phase.
We have also studied the effect of an applied electric
field that shows a complex 
behavior depending upon the temperature.

\section{Experimental Procedure}

The liquid crystal used in our studies, KN125, was synthesized
using established procedures \cite{ratna93}.
The chemical structure of this molecule is shown in Fig.\ 2.
This material exhibits a large electroclinic effect 
over a wide temperature range \cite{crawford94}.
The phase sequence for KN125 upon heating is
Crystal -- 29 $^\circ$C -- Smectic-$A$ -- 80 $^\circ$C -- Isotropic.
The smectic-$A$ phase can be supercooled below
25 $^\circ$C for several days without crystallizing.
CD studies were performed on a Jasco J-720 spectropolarimeter.
All data are given in units of molar circular dichroism
$\Delta \varepsilon \equiv \varepsilon_L - \varepsilon_R$,
where $\varepsilon_{L(R)}$ is the molar decadic absorption
coefficient for left- (right-) circularly polarized light
(e.g. the ratio of transmitted to incident intensity is
$I_t / I_i = 10^{- \varepsilon c d}$, where $c$ is the
molar concentration and $d$ is the pathlength in cm).

Studies in the planar geometry were performed using 
commercial liquid crystal cells (E.H.C. Co., Tokyo)
with a nominal cell gap of 2.5 $\mu$m.
These cells had a rubbed polyimide surface layer
and were mounted in an Instec MK1 hot stage.
The cells were oriented so that the sample
optic axis was rotated by 45 deg. from the optic axis of the
polarization modulator in the spectrometer to eliminate
artifacts from the inherent birefringence of the 
sample (see discussion below).
A 100 Hz bipolar square wave of variable amplitude was
applied to 1 cm$^2$ ITO electrodes.
All fields are given in RMS voltage.
For probing the CD along the layer normal 
direction, a cell having a 12 $\mu$m gap and
treated with octadecyltrichlorosilane to promote homeotropic
alignment (molecules perpendicular to the surface) was used.

\section{Results}

The CD spectra of KN125 measured in two geometries are
shown in Figure 2(a).
These spectra were both taken at 25 $^\circ$C with no applied field.
It was ascertained using optical microscopy that the sample
had a uniform optic axis and was in the smectic-$A$ phase, with no
signs of crystallization.
The spectra in Fig.\ 2(a) reveal a large peak around 362 nm
when the molecules are aligned in the planar geometry (solid line),
while such a peak is not seen for homeotropic alignment (dashed line).
The small wiggles at longer wavelengths are interference effects 
due to the finite thickness of the sample.
Two precursor compounds were also studied to establish
the origin of the peak in the CD spectra.
The first precursor, denoted HNBB, consisted of the
core of KN125 without the acyl chains, and the second, HBB,
consisted of KN125 without either the chains or
the nitro substitution.
Absorption spectra of these compounds in solution, along
with their chemical structures, 
are shown in Fig.\ 2(b).
A broad peak centered near 360 nm appears in the
HNBB spectra (solid line)
and is absent in the HBB spectra (dashed line).
This shows that the peak in the CD spectra at 362 nm
reflects an electronic transition related to the laterally
substituted nitro group.
The wavelength indicates that the $n_a \rightarrow \pi^\ast$
transition is slightly red-shifted due to coupling of the nitro
electronic orbitals to those in the biphenyl group \cite{jaffe62}.
It should be pointed out that the CD peak shown
in Fig.\ 2 is more than 2000 times larger than the intrinsic
CD of KN125 in solution ($\Delta \varepsilon_{355} = -0.035$
cm$^{-1}$ M$^{-1}$ in acetonitrile).
Comparison of the CD and UV spectra gives a dissymmetry factor
$g \equiv \Delta \varepsilon / \varepsilon = -0.36$.

An important test of chiral phenomena is to study
the CD spectra from opposite enantiomers,
since any effect which arises purely from the
handedness of the molecule, should change sign,
but not magnitude, when the chirality of the
molecule is reversed.
In Figure 3 we show the CD spectra
from both enantiomers of KN125 at 25 $^\circ$C
in planar cells.
We clearly see that reversing the molecular
chirality leads to a concurrent change in the sign of
the CD peak, but no change in its magnitude.

Although this large CD signal implies a twist of the
chiral molecules, we have to carefully eliminate
all possible experimental 
artifacts which could be contributing to the CD signal.
One obvious problem can be caused by a misalignment
of the rubbing directions at the two surfaces 
leading to a small twist through the cell.
We have performed experiments using numerous cells, both 
commercially obtained and fabricated in-house,
taking care in each case to see that the rubbing directions 
at the two surfaces are accurately antiparallel (within 1$^\circ$).
We find the peak magnitude varies by less than 10\% 
between cells under similar conditions
(temperature, field, alignment layer).
More importantly, we always observe a negative CD peak
from the {\it (R)}-enantiomer of KN125, while
the {\it (S)}-enantiomer always gives a positive peak.

Another important factor to consider is the effect of linear dichroism.
A number of authors have discussed 
the difficulties in deconvoluting the linear and circular
contributions to signals obtained using 
polarization-modulation spectroscopy \cite{davidsson80,kuball,shindo}.
In particular, Shindo and coworkers \cite{shindo} have elegantly
addressed this problem using the Stokes-Mueller formalism \cite{go66}.
They find that the CD signal measured 
in a polarization-modulation spectrometer is given by
\begin{eqnarray}
\Delta \varepsilon_{\rm meas} & = &
\Delta \varepsilon_{\rm true} + 
\frac{1}{2} \left( \mbox{LD}^\prime ~ \mbox{LB}
{}~ - ~ \mbox{LD} ~ \mbox{LB}^\prime \right) \nonumber \\
& & + \left(\mbox{LD}^\prime \sin 2 \theta
{}~ - ~ \mbox{LD} \cos 2 \theta \right)
\sin \alpha
\label{cdeqn}
\end{eqnarray}
where LD (LB) is the linear dichroism (birefringence) with respect
to the optic axis of the material,
LD$^\prime$ (LB$^\prime$) is the linear dichroism
(birefringence) with respect
to an axis rotated 45 degrees from the optic axis around the
direction of light propagation, $\theta$ is the angle
between the optic axis of the sample and that of the
spectrometer, as defined by its optical modulator, and $\alpha$
is an angle representing any imperfections in the
optical modulator due to misalignment 
or residual birefringence.
Note that in these units
$\mbox{LD} = \varepsilon_\parallel - \varepsilon_\perp$ and
$\mbox{LB} = 4 \pi ( n_\parallel - n_\perp ) /
( c \, \lambda_0 \ln 10 )$,
where $c$ is the molar concentration, $\lambda_0$ is the
vacuum wavelength of the incident light, and
$n_{\parallel (\perp)}$ is the index of refraction
parallel (perpendicular) to the optic axis.
Similar relations hold for LD$^\prime$ and LB$^\prime$.
The first component of the measured signal is the true
CD signal which is independent of $\theta$.
The second term is a combination of the linear dichroism
and linear birefringence as measured both along 
the optic axis of the sample and at 45 degrees to it.
It is also rotationally independent and is only
nonzero for a biaxial sample where
LD$^\prime$, LB$^\prime \neq 0$.
Since we find the linear dichroism at 45 degrees to
be at least $10^3$ times smaller than that measured 
along the optic axis,
(LD$^\prime$/LD $\alt 10^{-3}$),
our sample is approximately uniaxial and we
ignore the second term.
The final term in Eqn.\ (\ref{cdeqn}) is a rotationally
dependent term due to the interaction between a nonideal
optical modulator and a sample with linear dichroism.
Indeed, we find that the measured CD signal has 
a small rotationally dependent component which does
not change sign
when the chirality of the molecule is switched \cite{spector}.
The angular dependence of this component is consistent
with a small misalignment of the 
photoelastic modulator ($\alpha = 1.1^\circ$).

Shindo {\it et al.} also consider additional artifacts due to 
a partially polarizing detector and imperfect harmonic response
of a lock-in amplifier \cite{shindo}.
A quartz depolarizer inserted before the detector
produces no change in the measured CD 
ruling out the former.
The latter is very difficult to quantify.
However, no evidence of such artifacts is seen
when an achiral, linearly anisotropic sample is studied.
When the sample is aligned so that the optic axis of the liquid 
crystal is 45 degrees with respect to the modulator,
the contribution to the CD due to linear effects is 
expected to vanish in a uniaxial sample.
All measurements discussed in this paper
were made in such a geometry.

Having thus eliminated experimental artifacts,
we shall now discuss the effect of temperature and applied 
electric field on the CD spectra.
Figure 4 shows the temperature variation of the CD peak 
in the absence of an applied electric field.
CD spectra at five different temperatures are
shown in Fig.\ 4(a) for {\it (R)}-KN125.
As the temperature increases, the magnitude of the
CD peak continually decreases, but 
its position remains unchanged.
This confirms that the observed effect is due
to preferential absorption of right-circularly
polarized light,
as opposed to a preferential scattering effect,
where the wavelength of the CD 
peak is expected to shift with temperature.
Figure 4(b) shows the temperature dependence of the CD peak 
magnitude, $\Delta \varepsilon_{pk}$, for both enantiomers
of KN125 at 25 $^\circ$C with no applied field.
We see that both enantiomers show a 
similar decrease in $\Delta \varepsilon_{pk}$
with increasing temperature.
However, the CD peak does not continually vanish as
the isotropic transition is approached.
Rather, there remains a small remnant 
value over a wide temperature range in the
smectic-$A$ phase right up to the melting point.

When an AC electric field is applied across the sample
at 25 $^\circ$C, we also find a change in the magnitude of the 
CD peak, but not in its position.
As the field is increased, the CD peak magnitude continually decreases, 
as shown in Figure 5.
Both enantiomers show a similar field dependence of
the peak CD.  
However, the CD peak does not completely disappear at high fields.
Instead, we find a small residual CD 
signal at high field
($\Delta \varepsilon_{pk}$ = -14.7 at 25 $^\circ$C, 5.2 V/$\mu$m)
which is similar to that seen in Fig.\ 4(b) at higher 
temperatures and no field
($\Delta \varepsilon_{pk}$ = -14.9 at 75 $^\circ$C, 0 V/$\mu$m).
This remnant CD indicates the existence of 
a small twist which cannot be unwound by electric field
or thermal effects.

As the temperature is increased in the smectic-$A$ phase,
the situation becomes more complex.
The field dependence of the peak magnitude is shown
for both enantiomers at 40 and 60 $^\circ$C in Fig.\ 6(a).
Near 40 $^\circ$C, the peak magnitude is relatively field
independent (open symbols).
At higher temperatures, $\Delta \varepsilon_{pk}$
begins to increase with field, as seen in the data
at 60 $^\circ$C (filled symbols).
The temperature dependence of the peak circular
dichroism of {\it (R)}-KN125 is shown for three different
fields in Fig.\ 6(b).
We find significant differences in the temperature dependence
of the CD peak at different fields.
At zero field, the peak decreases monotonically as the temperature
is increased as also shown in Fig.\ 4(b).
At an intermediate field of 2.6 V/$\mu$m,
the peak CD is relatively temperature independent,
while at large field, 5.2 V/$\mu$m, the peak initially increases with
temperature to a maximum value around 60 $^\circ$C and then 
begins to decrease.
At all fields the peak goes to zero at the isotropic phase transition.
Similar behavior is also seen in the {\it (S)}-enantiomer.
This complex dependence of the CD on both temperature and field 
provides further evidence that the observed results are not
artifacts due to linear effects, since 
measurements of the linear dichroism along the optic axis
reveal that it is weakly dependent on both 
temperature and an applied AC field \cite{spector}.

\section{Discussion}

A number of possible explanations may describe the origin
of the large CD peak observed in the planar, 
but not homeotropic, geometry.
The possibility of experimental artifacts was discussed in detail above.
Another possible explanation is in-plane chiral fluctuations
of the type proposed by 
Lubensky, Kamien, and Stark \cite{lubensky96}.
While the smectic-$A$ phase has 
no macroscopic chiral order, it is possible for the molecular chirality
to bias the thermal fluctuations of the molecules giving the
system local, or short-range, chiral order.
However, since a smectic-$A$ cannot 
support long-range twist deformations \cite{degennes},
these fluctuations can only extend over a few molecules and 
are unlikely to be responsible for the large
enhancement we see in the CD spectra.

Instead, we believe the large chiral effect is due
to a twisting of the director induced by the surface 
electroclinic effect discussed earlier.
Here, the polar interactions at the cell wall-liquid crystal interface
act like an applied field, causing the molecules to be
tilted with respect to the layer normal by an angle $\Psi$.
Since the molecules are strongly anchored by the rubbing,
the layer normal is then tilted by $\Psi$ from 
the rubbing direction as depicted in Fig.\ 1.
The direction of the layer normal established
by the surface propagates unchanged into the bulk.
The layer normals induced by opposing cell
interfaces are rotated in 
opposite directions since the interfaces are
mirror images of each other.
This implies that the molecules in the
two interfacial regions twist in the same
direction so that
the net twist induced by the two cell walls
is twice that from a single interface.
However, this also implies that the layer
normal rotates by $2\Psi$ between the two surfaces.
Since uniform twisting of the layer normal is energetically
disfavored in the smectic-$A$ phase \cite{degennes},
this rotation presumably occurs over a very small
distance in a twist grain boundary inside the cell \cite{renn88}.
Patel, Lee, and Goodby observed such defects in thin 
cells, but not in thicker cells where
only one domain was seen \cite{patel91}.
In the thick cells, the observation of a single
domain with optic axis not parallel to the rubbing direction
was attributed to preferential growth of
one of the two possible layer normal directions
into the bulk \cite{patel91}.
We observe good optical extinction when the cell is viewed between
crossed polarizers with one polarizer aligned 
along the optic axis and find the optic axis of the sample is
rotated by about 10$^\circ$ from the rubbing direction.
Interestingly, this 10$^\circ$ offset is similar to the zero
field mosaicity measured in x-ray diffraction 
experiments on the same material \cite{geer98}.
We find that the offset angle changes by less than a
degree over the entire smectic-$A$ phase.

Further support for the surface electroclinic effect model
is obtained by studying a cell with single-sided alignment.
Recent work has shown that KN125 can be well aligned in a cell
where only one side has rubbed polyimide on it,
the other surface being plain conducting (ITO) glass \cite{hermanns98}.
The thickness of the cell was approximately the same as the
double-sided alignment layer cell described earlier.
Fig.\ 3 shows the CD spectra from such a single-sided alignment
layer cell containing {\it (R)}-KN125. 
We find that the CD peak is decreased compared to the
double-sided cell, but still quite large.
In fact the peak magnitude is slightly larger than half 
that from the double-sided cell, possibly indicating a reduction 
in twist at the grain boundary.
This observation rules out chiral fluctuations in the bulk
as a major contributor to the observed CD, since 
such fluctuations should only depend on the thickness
of the cell and not on whether one or both sides have 
an alignment layer.
We also find the temperature and field dependence of the
CD peak in the singled-sided alignment layer cell follows
that of the double-sided cell, as shown in Figs.\ 4(b) and 5,
respectively (diamonds).

In order to explain the complex field dependence of
the CD spectra with increasing temperature, we 
consider two temperature regions.
At low temperature ($\sim$ 25-35 $^\circ$C),
the CD peak decreases rapidly with applied field,
while at high temperature ($\sim$ 50-80 $^\circ$C),
the CD increases with field.
In the low temperature region, the material responds as
though it were near the smectic-$A$ to smectic-$C^\ast$ transition.
This is also seen in the optical response of KN125,
where the electroclinic coefficient $e_c$ is found to 
increase sharply with decreasing temperature as shown
in Fig.\ 4(b) (solid circles).
The divergence of $e_c$ indicates an apparent
transition to a tilted smectic phase at $\sim 20~^\circ$C,
whose occurrence is preempted by crystallization.

Similar pretransitional behavior is also expected to affect the surface
properties of the liquid crystal.
Near the apparent bulk transition, the
surface region should freeze into an
ordered phase --- either a tilted, chiral
smectic phase, with a twist of the molecular
director from layer to layer, or the
chiral-stripe phase discussed in Ref.\ \cite{selinger},
with a twist of the
molecular director in the smectic layer plane.
In either case, the chiral
environment of the director should enhance the CD signal compared with the
uniform smectic-$A$ phase.
As the temperature decreases, the penetration depth of the
ordered phase into the interior should increase,
and hence the CD signal should increase.
Indeed, Chen {\it et al.} confirmed that the
size of the interfacial region diverges as the
smectic-$C^\ast$ transition temperature
is approached \cite{chen92}.
This is consistent with the result shown in Fig.\ 4(b) that
$|\Delta\varepsilon_{pk}|$ increases sharply as
the temperature decreases near
the apparent transition.
Furthermore, when a voltage is applied across the
cell, the twist of the director in the ordered phase should be at least
partially suppressed, leading to a decrease in the CD signal \cite{comment}.
This agrees with the observation in Fig.\ 5 that
$|\Delta\varepsilon_{pk}|$ decreases as the electric
field increases in this temperature range.
However, the twist in the interfacial region cannot be
unwound by the field, leaving a remnant CD at high field.

At high temperature, the behavior should be quite different.  In this
temperature regime, surface freezing becomes unimportant, and the liquid
crystal is purely in the smectic-A phase.
For that reason, the chiral twisting
of the director should be dominated by the surface electroclinic effect.
To find the profile of the electroclinic tilt
angle near a surface, we must
minimize the free energy subject to the appropriate boundary conditions.
The free energy can be written as
\begin{eqnarray}
F &&=\int d^3 x \left[\textstyle\frac{1}{2} K_2 |\nabla\theta|^2
-\textstyle\frac{1}{2}a({\bf n}\cdot{\bf N})^2
- s({\bf E}\cdot{\bf n}\times{\bf N})({\bf n}\cdot{\bf N}) \right]\nonumber\\
&&=\int d^3 x \left[\textstyle\frac{1}{2} K_2 |\nabla\theta|^2
-\textstyle\frac{1}{2}a\cos^2\theta - sE\sin\theta\cos\theta \right],
\end{eqnarray}
where $\bf n$ is the molecular director, $\bf N$ is the 
layer normal, $\theta$ is the tilt angle between $\bf n$
and $\bf N$, and $K_2$ is the twist elastic
constant.  Note that this free energy is invariant under the
symmetries ${\bf n}\to-{\bf n}$ and ${\bf N}\to-{\bf N}$.
The final term represents the bulk electroclinic coupling of
the tilt angle to the field, with
$s$ known as the structure coefficient \cite{andersson88}.
In the limit of small
$\theta$, this free energy reduces to the more familiar expression
$F=\int d^3 x [\frac{1}{2} K_2 |\nabla\theta|^2 + \frac{1}{2}a\theta^2 -
sE\theta]$, plus a constant.

To minimize the free energy, we rewrite it in the form
\begin{equation}
F =\int d^3 x \left[\textstyle\frac{1}{2} K_2 |\nabla\theta|^2
-\textstyle\frac{1}{4}a'\cos2(\theta-\theta_0)\right],
\end{equation}
plus an irrelevant constant, where
\begin{mathletters}
\begin{eqnarray}
&& a'=[a^2 + (2sE)^2]^{1/2},\\
&& \theta_0=\frac{1}{2}\arctan\frac{2sE}{a}.
\end{eqnarray}
\end{mathletters}
In the bulk, the minimum of the free energy is $\theta=\theta_0$; i.e.
$\theta_0$ is the bulk electroclinic tilt.
Near a surface defined by $y=0$, the minimum of the
free energy is given by the Euler-Lagrange equation
\begin{equation}
-K_2 \frac{d^2\theta}{dy^2} + \frac{1}{2}a'\sin2(\theta-\theta_0) = 0 .
\end{equation}
This differential equation is a standard sine-Gordon soliton equation.
Because of the surface electroclinic effect, we impose
the boundary condition
$\theta=\Psi$ at $y=0$, where $\Psi$ is the surface tilt.
In the bulk, the
boundary condition is $d\theta/dy\to0$ as $y\to\infty$.
Solving the Euler-Lagrange equation with these boundary
conditions gives the tilt profile
\begin{equation}
\theta(y) = \theta_0 + 2\arctan\left[\left(\tan\frac{\Psi-\theta_0}{2}\right)
e^{-y/\xi}\right],
\label{tilt profile}
\end{equation}
where $\xi=(K_2/a')^{1/2}$ is the tilt correlation length.

In principle, one might want to calculate the CD
signal from the tilt profile $\theta(y)$.
In practice, however, the CD signal is a complex molecular
property that depends on the detailed interactions
of the liquid crystal with light.
Even without a theory for the optical properties of the molecules,
we can say that the CD signal is a chiral optical
property that depends on the
chiral parameters of the system. 
 From the tilt profile, we can determine how typical
chiral parameters depend on the applied electric field.
The simplest chiral parameter that one might
calculate is the integrated twist
\begin{equation}
Q_1(E) = \int_{0}^{\infty} dy \, \frac{d\theta}{dy}.
\end{equation}
This integrated twist is fixed by the boundary conditions to be
$Q_1(E) = \theta_0-\Psi$.
Because the experiment averages over the front and
back surfaces of the cell, and over the full period
of an AC electric field, it is
appropriate to calculate the average chiral parameter
\begin{equation}
\overline{Q_1}(E)=\frac{Q_1(E)+Q_1(-E)}{2} = -\Psi.
\end{equation}
Note that this chiral parameter is independent of $E$.  However,
there are many measures of chirality which can correlate
with the CD signal \cite{harris99}.
One possibility, allowed by symmetry, is
\begin{equation}
Q_3(E) = \int_{0}^{\infty} dy \left(\frac{d\theta}{dy}\right)^3,
\end{equation}
which is not fixed by the boundary conditions. 
 From (\ref{tilt profile})
we find the field-averaged result
\begin{equation}
\overline{Q_3}(E)=\frac{Q_3(E)+Q_3(-E)}{2}
= \frac{1}{4\xi^2}\left[- 2\Psi + \sin2\Psi\cos2\theta_0\right].
\end{equation}
This result shows that the average chiral
parameter $\overline{Q_3}(E)$ is nonzero only when
the system has a nonzero surface electroclinic tilt $\Psi$.
This average chiral parameter is nonzero at $E=0$ and grows
larger in magnitude as $E$ increases.
For that reason, in the high-temperature regime, we expect
that an applied field should also give an increase in the CD signal.
This is consistent with the experimental results of Fig.\ 6.
Furthermore, we can estimate how much the applied field
should increase the CD signal.
The result above implies that
\begin{equation}
\frac{\overline{Q_3}(E)-\overline{Q_3}(0)}{\overline{Q_3}(0)}
= \frac{\frac{1}{\cos2\theta_0} - 1}{1 - \frac{\sin2\Psi}{2\Psi}}
\approx \alpha E^2,
\end{equation}
with $\alpha=3 e_c^2/\Psi^2$, where
$e_c \equiv d\theta_0/dE |_{E \rightarrow 0} = s / a$ is the bulk
electroclinic coefficient, and the approximation holds for
small $\theta_0$ and $\Psi$.
We have $\Psi\approx 10$~deg at all temperatures
and $e_c = 0.78$~deg~(V/$\mu$m)$^{-1}$ at $60~^\circ$C,
and hence we expect $\alpha\approx 0.018$~(V/$\mu$m)$^{-2}$.
By comparison, the solid line in
Fig.\ 6(a) shows a fit of the $60~^\circ$C data to the parabolic form
\begin{equation}
\Delta\varepsilon_{pk}(E) = \Delta\varepsilon_{pk}(0)
\left( 1 + \alpha E^2 \right) ,
\label{cdvsfield}
\end{equation}
which gives the coefficient $\alpha = 0.046$~(V/$\mu$m)$^{-2}$.
Thus, our simple model gives at least the
right order of magnitude for the sensitivity
of the CD signal to applied electric field.

Of course, $\overline{Q_3}(E)$ is 
not the only pseudoscalar measure of chirality.  While
$\overline{Q_1}(E)$ was independent of $E$, we could consider
\begin{equation}
Q_1(E;Y) = \int_0^Y dy \, {d\theta\over dy}.
\end{equation}
where $Y$ would be the lengthscale over which surface effects  
modify the smectic order leading to an enhanced CD signal.
Because of the nonlinear dependence of $\theta(y)$ on $E$ in
Eq.\ (\ref{tilt profile}), the chiral parameter $\overline{Q_1}(E;Y)$
will depend quadratically on $E$.
Moreover, if $Y\sim\xi$, the coefficient of
the quadratic term will have the same order of magnitude as $\alpha$
calculated above.
This will be a generic feature of any chiral parameter
constructed from $\theta(y)$ -- the quadratic coefficient
will always be of order $e_c^2/\Psi^2$, in agreement
with the experimental result.

In conclusion, molecular twisting within the layer
of a chiral smectic-$A$ has been studied using circular
dichroism spectroscopy.
The results have been discussed in terms
of a model based on the surface electroclinic effect.
At high temperatures, where the bulk is purely
smectic-$A$, an applied field leads to an
increase in the chiral parameter $Q_3$ which can
account for the observed field dependence of the
CD spectra.
At lower temperatures, surface effects 
enhance the chiral, pretransitional behavior.

\section{Acknowledgments}

We thank Jawad Naciri and Anna Davis for synthesis of
the samples used in our studies and T. C. Lubensky,
H.-G. Kuball, and S. Chandrasekhar
for helpful discussions.
This work was supported by the Office of Naval Research.
RDK was supported in part by NSF CAREER Grant DMR97-32963.

\begin{figure}
\caption{Schematic representation of the
surface electroclinic effect.
At the surface, the molecular director {\bf n}
lies along the rubbing direction, while the
layer normal {\bf N} deviates by an angle $\Psi$.
In the boundary layer of thickness $\xi$, the
director twists from the rubbing axis until
it reaches the layer normal direction, while
{\bf N} remains constant.}
\end{figure}

\begin{figure}
\caption{(a) Circular dichroism of KN125 in planar
geometry (solid line) and homeotropic geometry
(dashed line).  Both spectra were recorded
at 25 $^\circ$C on the {\it (R)}-enantiomer
with no applied field.
The chemical structure of KN125
is shown with the
chiral center indicated by $\ast$.
(b)  Absorption spectra of precursor
compounds HNBB (solid line) and HBB
(dashed line) in solution.
Also shown are the chemical structure of
HNBB, the core of KN125 without
the acyl chains, and
HBB, the same molecule without the
laterally substituted nitro group.}
\end{figure}

\begin{figure}
\caption{Circular dichroism from opposite enanatiomers
of KN125 in at 25 $^\circ$C.
The solid lines show the spectra from KN125 in cells
with an alignment layer on both sides,
while the dashed line shows {\it (R)}-KN125 in a
single-sided alignment layer cell.}
\end{figure}

\begin{figure}
\caption{Temperature dependence of the
circular dichroism of KN125.
(a) The CD spectrum of the
{\it (R)}-KN125 at different temperatures in
the absence of an applied field.
As the temperature increases, the magnitude
of the peak decreases and becomes unobservable
above the melting temperature.
The position of the peak does not change with temperature.
(b) The magnitude of the CD peak versus temperature
for  {\it (R)}-KN125 ($\circ$)
and  {\it (S)}-KN125 ($\Box$) in cells
with double-sided alignment and 
for  {\it (R)}-KN125 in a single-sided
alignment cell ($\Diamond$).
Also shown is the temperature dependence of the
electroclinic coefficient ($\bullet$).}
\end{figure}

\begin{figure}
\caption{Field dependence of the circular
dichroism of KN125 at 25 $^\circ$C.
The magnitude of the CD peak versus applied
electric field is shown
for {\it (R)}-KN125 ($\circ$)
and {\it (S)}-KN125 ($\Box$) in cells
with double-sided alignment and 
for {\it (R)}-KN125 in a single-sided
alignment cell ($\Diamond$).}
\end{figure}

\begin{figure}
\caption{Field dependence of the circular dichroism
of KN125 fields at higher temperatures.
(a) The magnitude of the
CD peak versus electric field for
for {\it (R)}-KN125 ($\circ$)
and {\it (S)}-KN125 ($\Box$).
The open symbols represent data at 40 $^\circ$C,
while the filled symbols are at 60 $^\circ$C.
The dashed lines are guides to the eye, while
the solid line represents a fit of the
{\it (R)}-KN125 data at 60 $^\circ$C to the
quadratic behavior of Eqn.\ (\protect\ref{cdvsfield}).
(b) Temperature dependence of the magnitude
of the CD peak of {\it (R)}-KN125 ($\circ$)
at applied fields of 0.0 V/$\mu$m ($\circ$),
2.6 V/$\mu$m ($\Box$), and 5.2 V/$\mu$m ($\Diamond$).}
\end{figure}


\begin{references}
\bibitem{comment} Reference~\protect\cite{selinger} shows
that an electric field can lead to an increase in
the in-plane molecular twist in the direction
orthogonal to the applied field.
However, the CD experiment should be
sensitive to a molecular twist across the cell,
in the direction parallel to the applied field.
This twist in this direction is suppressed by the field.
\end{references}
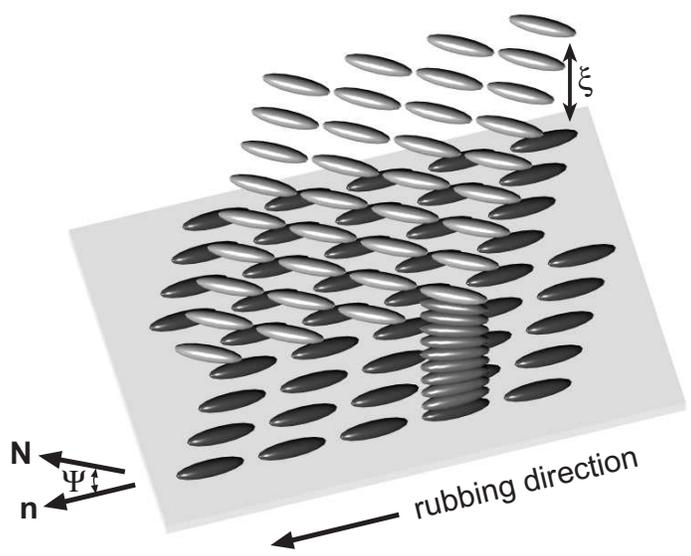
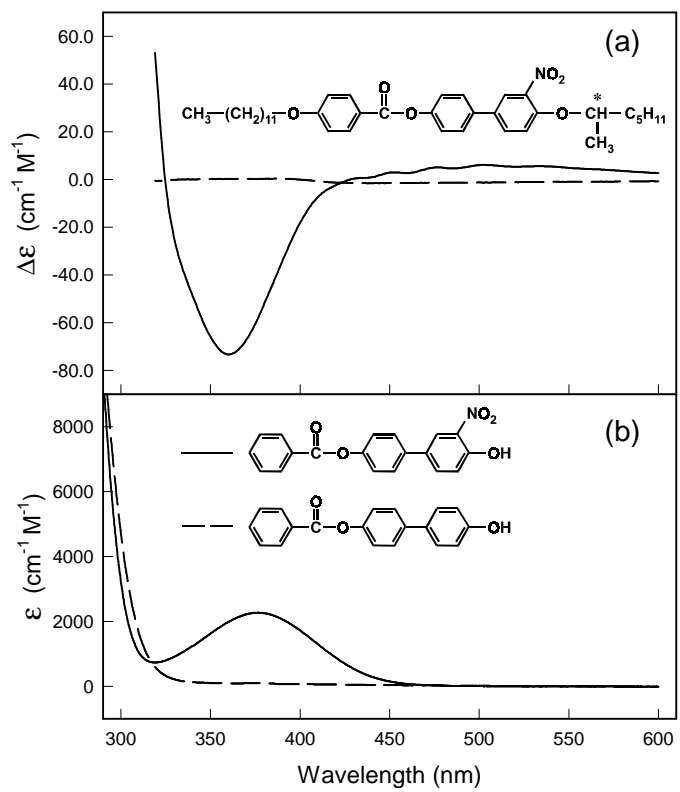
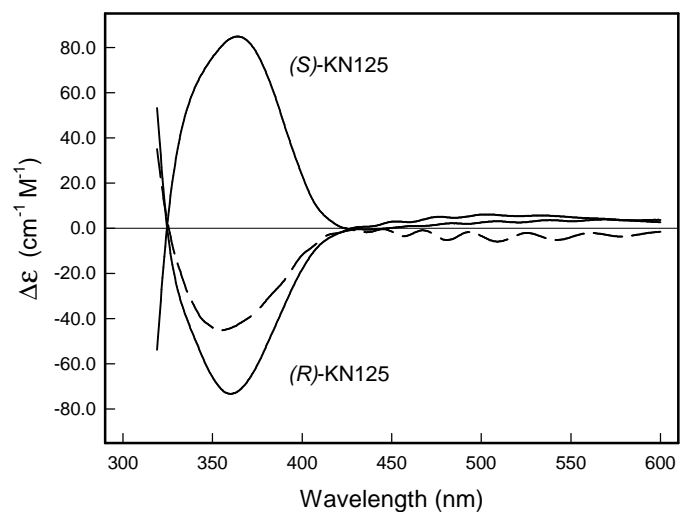
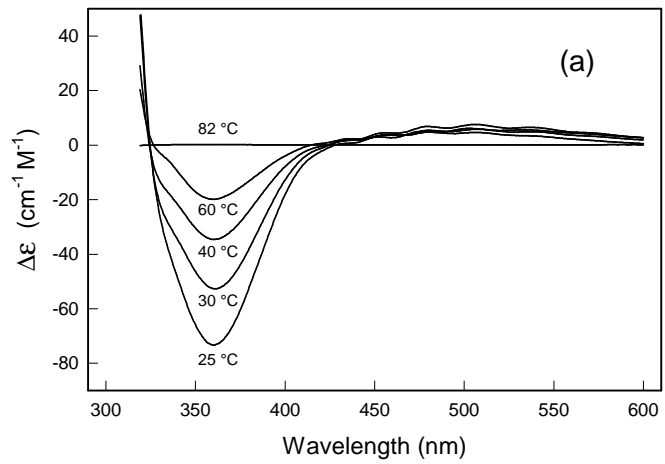
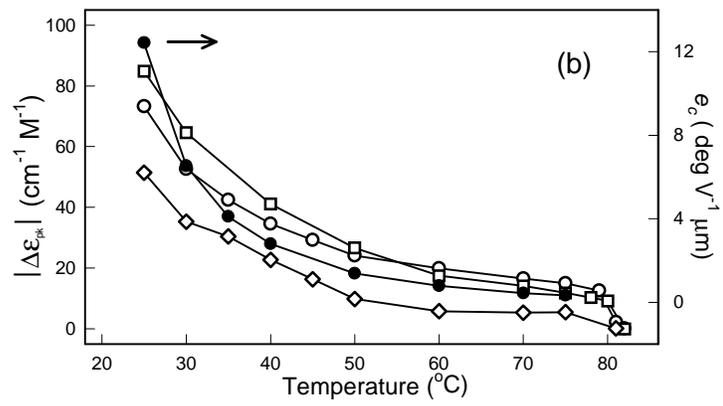
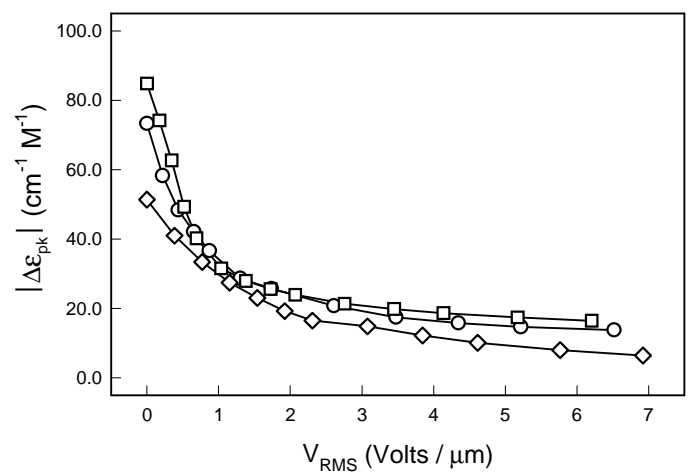
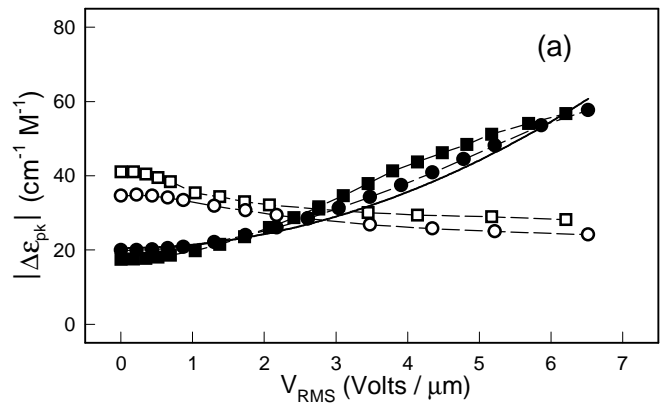
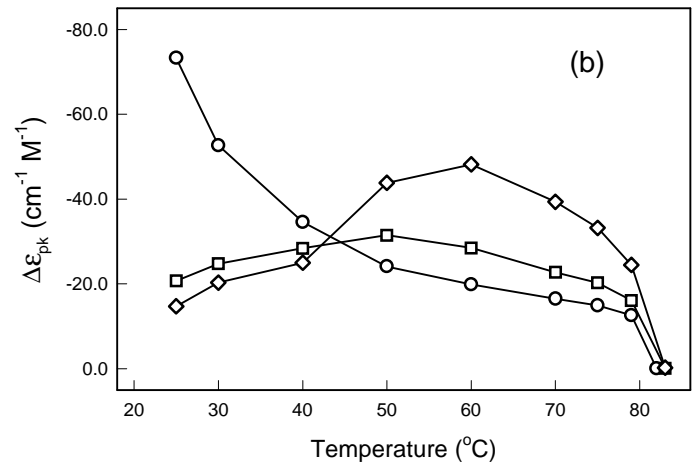
\end{document}